\documentclass{ptephy_v1}

\preprintnumber{XXXX-XXXX} 

\usepackage{amsmath,cases}
\usepackage{lastpage}
\usepackage{graphicx}
\usepackage{epstopdf}
\usepackage{mathtools}
\usepackage{booktabs}

\usepackage{braket}
\usepackage{comment}
\usepackage{scalerel}
\usepackage{tikz}
\usepackage[hang,small,bf]{caption}
\usepackage[subrefformat=parens]{subcaption}
\usepackage{multirow}
\usepackage{lineno}
\usetikzlibrary{svg.path}
\newcommand{\Var}{\mathrm{Var}}

\nolinenumbers
\definecolor{orcidlogocol}{HTML}{A6CE39}
\tikzset{
	orcidlogo/.pic={
		\fill[orcidlogocol] svg{M256,128c0,70.7-57.3,128-128,128C57.3,256,0,198.7,0,128C0,57.3,57.3,0,128,0C198.7,0,256,57.3,256,128z};
		\fill[white] svg{M86.3,186.2H70.9V79.1h15.4v48.4V186.2z}
		svg{M108.9,79.1h41.6c39.6,0,57,28.3,57,53.6c0,27.5-21.5,53.6-56.8,53.6h-41.8V79.1z M124.3,172.4h24.5c34.9,0,42.9-26.5,42.9-39.7c0-21.5-13.7-39.7-43.7-39.7h-23.7V172.4z}
		svg{M88.7,56.8c0,5.5-4.5,10.1-10.1,10.1c-5.6,0-10.1-4.6-10.1-10.1c0-5.6,4.5-10.1,10.1-10.1C84.2,46.7,88.7,51.3,88.7,56.8z};
	}
}
\newcommand\orcidicon[1]{\href{https://orcid.org/#1}{\mbox{\scalerel*{
				\begin{tikzpicture}[yscale=-1,transform shape]
				\pic{orcidlogo};
				\end{tikzpicture}
			}{|}}}}
\usepackage{hyperref}
		
\begin{document}
\title{Simultaneous determination of the cosmic birefringence and miscalibrated polarisation angles from CMB experiments}
\author[1,*]{Yuto Minami}
\affil[1]{High Energy Accelerator Research Organization, 1-1 Oho, Tsukuba, Ibaraki 305-0801, Japan \email{yminami@post.kek.jp}}

\author[2]{Hiroki Ochi}
\affil[2]{Graduate School of Engineering Science, Yokohama National University, 79-5 Tokiwadai, Hodogaya-ku, Yokohama 240-8501, Japan}
\author[3,4]{Kiyotomo Ichiki}
\affil[3]{Graduate School of Science, Division of Particle and Astrophysical Science, Nagoya University, Chikusa-ku, Nagoya 464-8602, Japan}
\affil[4]{Kobayashi-Maskawa Institute for the Origin of Particles and the Universe, Nagoya University, Chikusa-ku, Nagoya 464-8602, Japan}
\author[5]{Nobuhiko Katayama}
\affil[5]{Kavli Institute for the Physics and Mathematics of the Universe (Kavli IPMU, WPI), Todai Institutes for Advanced Study, The University of Tokyo, Kashiwa 277-8583, Japan}
\author[5,6]{Eiichiro Komatsu}
\affil[6]{Max Planck Institute for Astrophysics, Karl-Schwarzschild-Str. 1, D-85748 Garching, Germany}
\author[5]{Tomotake Matsumura}
\begin{abstract}
We show that the cosmic birefringence and miscalibrated polarisation angles can be determined simultaneously by cosmic microwave background (CMB) experiments using the cross-correlation between $E$- and $B$-mode polarisation data. This is possible because polarisation angles of the CMB are rotated by both the cosmic birefringence and miscalibration effects, whereas those of the Galactic foreground emission only by the latter. Our method does not require prior knowledge of the $E$- and $B$-mode power spectra of the foreground emission, but uses only the knowledge of the CMB polarisation spectra. Specifically, we relate the observed $EB$ correlation to the difference between the \textit{observed} $E$- and $B$-mode spectra in the sky, and use different multipole dependence of the CMB (given by theory) and foreground spectra (with no assumption) to derive the likelihood for the miscalibration angle $\alpha$ and the birefringence angle $\beta$. We show that a future satellite mission similar to LiteBIRD can determine $\beta$ with a precision of ten arcminutes.
\end{abstract}
\subjectindex{xxxx, xxx}

\maketitle
\section{Introduction}\label{sec:Introduction}
Cross-correlation between $E$- and $B$-mode polarisation of the cosmic microwave background (CMB) is sensitive to parity-violating physics in the Universe.
One of the physical effects, known as  ``cosmic birefringence''~\cite{Carroll:1998zi,Lue:1998mq,Feng:2004mq,Feng:2006dp},
rotates CMB polarisation angles as CMB photons propagate to us since last scattering at $z\approx 1100$ via,
e.g., a Chern-Simons coupling between a light scalar field and the electromagnetic tensor~\cite{Carroll:1998zi}.

However, this effect is degenerate with an artificial rotation of polarisation angles by miscalibration of the orientation of the instrument~\cite{Komatsu:2010fb}.
If we use the observed cosmological $EB$ correlation to solve for the miscalibration angle $\alpha$~\cite{Keating:2012ge},
we may run into two issues:
(1) we lose sensitivity to the cosmic birefringence angle $\beta$;
and
(2) we would infer a wrong value of $\alpha$ if there were non-vanishing $\beta$.
In this paper, we mitigate this issue by using the polarised Galactic foreground emission.
Polarisation of the Galactic foreground is insensitive to the cosmic birefringence effect because of a limited propagation length of photons.
We can distinguish between $\alpha$ (affecting both CMB and foreground) and $\beta$ (affecting only CMB) using the different multipole dependence of the CMB and Galactic foreground polarisation power spectra.
This is possible because the observed $EB$ correlation is related to the difference between the \textit{observed} $E$- and $B$-mode power spectra of the sky (including CMB and foreground)~\cite{Zhao:2015mqa},
and we know the CMB power spectra well;
thus, we can simultaneously determine $\alpha$ and $\beta$ without prior knowledge of the foreground polarisation power spectra.

Throughout this paper,
we shall assume that $\beta$ is uniform over the sky,
as this is the case that is degenerate with $\alpha$. There are physical mechanisms to produce spatially-varying $\beta$ \cite{Kosowsky:1996yc,Gardner:2006za,Pospelov:2008gg,Li:2008tma,Caldwell:2011pu},
which can be estimated from CMB polarisation data with a suitable estimator \cite{Kamionkowski:2008fp}.
See refs.~\cite{Ade:2015cao,Contreras:2017sgi,Array:2017rlf} for the current constraints on spatially-varying $\beta$.

The rest of the paper is organised as follows.
In Sect.~\ref{sec:methodology} we describe our methodology for evaluating the posterior distribution of $\alpha$ and $\beta$.
In Sect.~\ref{sec:results} we validate our method using sky simulations,
and present the main results.
We conclude in Sect.~\ref{sec:conclusions}.

\section{Methodology}\label{sec:methodology}
\subsection{Relating the observed $EB$ to the observed $EE-BB$}

When polarisation angles are rotated uniformly over the sky by an angle $\alpha$,
spherical harmonics coefficients of the observed $E$- and $B$-mode polarisation, denoted by ``o'', are related to the intrinsic ones by\footnote{
In this paper we do not discuss the $TB$ correlation but focus only on the $EB$ correlation, since a large cosmic variance in $T$ makes $TB$ less sensitive than $EB$ to $\alpha$ and $\beta$ when the instrumental noise is sufficiently low to measure $B$. In any case, it is straightforward to extend our method to include $TB$.
}
\begin{equation}\label{eq:EandBFromRotaiton}
\begin{split}
E_{\ell,m}^\mathrm{o} &= E_{\ell,m}\cos(2\alpha) 
- 
B_{\ell,m}\sin(2\alpha), \\
B_{\ell,m}^\mathrm{o} &= E_{\ell,m}\sin(2\alpha) 
+ 
B_{\ell,m}\cos(2\alpha).
\end{split}
\end{equation}
Throughout this paper, we shall adopt the notation that all spherical harmonics coefficients and power spectra have been multiplied by the appropriate beam transfer functions, unless noted otherwise. 

In this paper, we shall work with full-sky data without a mask,
as we do not wish to remove the Galactic foreground. 
Our method can be adopted straightforwardly to work in fractions of the sky (e.g., ground-based experiments), or with masks.

Defining the power spectra as $C_\ell^{XY}=(2\ell+1)^{-1}\sum_{m=-\ell}^\ell X_{\ell,m}Y_{\ell,m}^*$, we obtain \cite{Lue:1998mq,Feng:2004mq,Feng:2006dp}
\begin{align}
C_\ell^{EE,\mathrm{o}} &= C_\ell^{EE} \cos^2(2\alpha) + C_\ell^{BB} \sin^2(2\alpha) 
- C_\ell^{EB}\sin(4\alpha),
\label{eq:EEobsFromIntrinsic}
\\
C_\ell^{BB,\mathrm{o}} &= C_\ell^{EE} \sin^2(2\alpha) + C_\ell^{BB} \cos^2(2\alpha)
+ C_\ell^{EB}\sin(4\alpha),
\label{eq:BBobsFromIntrinsic}
\\
C_\ell^{EB,\mathrm{o}} &= \frac{1}{2} \left( C_\ell^{EE} -C_\ell^{BB} \right) \sin(4\alpha) +C_\ell^{EB}\cos(4\alpha).
\label{eq:EBobsFromIntrinsic}
\end{align}

Using Eq.~(\ref{eq:EEobsFromIntrinsic}) and (\ref{eq:BBobsFromIntrinsic}) in Eq.~(\ref{eq:EBobsFromIntrinsic}),
we find 
\begin{equation}\label{eq:EBandEEBB_obs}
C_\ell^{EB,\mathrm{o}} = \frac{1}{2}\left(  C_\ell^{EE,\mathrm{o}} -  C_\ell^{BB,\mathrm{o}} \right) \tan (4\alpha)+\frac{C_\ell^{EB}}{\cos(4\alpha)}\,.
\end{equation}
This result was first derived in ref.~\cite{Zhao:2015mqa} except for the $EB$ term. 
We can use Eq.~(\ref{eq:EBandEEBB_obs}) to solve for $\alpha$ \textit{with no assumption about the intrinsic} $C_\ell^{EE}-C_\ell^{BB}$. 
Then, we no longer have to worry about a bias
in $\alpha$ induced by incorrect modelling of the intrinsic $C_\ell^{EE}-C_\ell^{BB}$,
which was studied in ref.~\cite{Abitbol:2015epq}.

Next, we include the cosmic birefringence angle $\beta$ and noise (``N''), and write separately the foreground (``fg'') and CMB (``CMB'') components. We obtain
\begin{align}
E_{\ell,m}^\mathrm{o} &= 
E_{\ell,m}^\mathrm{fg}\cos(2\alpha)-  B_{\ell,m}^\mathrm{fg}\sin(2\alpha)
+E_{\ell,m}^\mathrm{CMB}\cos(2\alpha+2\beta)-  B_{\ell,m}^\mathrm{CMB}\sin(2\alpha+2\beta)
+E_{\ell,m}^\mathrm{N},
\\
B_{\ell,m}^\mathrm{o} &= 
E_{\ell,m}^\mathrm{fg}\sin(2\alpha) + B_{\ell,m}^\mathrm{fg}\cos(2\alpha)
+E_{\ell,m}^\mathrm{CMB}\sin(2\alpha+2\beta) + B_{\ell,m}^\mathrm{CMB}\cos(2\alpha+2\beta)
+B_{\ell,m}^\mathrm{N} .
\end{align}

From these coefficients,
we can relate $C_\ell^{EB,\mathrm{o}}$ to $C_\ell^{EE,\mathrm{o}}-C_\ell^{BB,\mathrm{o}}$ as 
\begin{equation}\label{eq:EBObsWithResiduals}
\begin{split}
\left( C_\ell^{EB,\mathrm{o}}\right.&
\left.
-\frac{\tan(4\alpha) }{2}
\left(C_\ell^{EE,\mathrm{o}}-C_\ell^{BB,\mathrm{o}}\right)
\right)\cos(4\alpha ) =
\\&
(C_\ell^{EE,\mathrm{CMB}} - C_\ell^{BB,\mathrm{CMB}})\sin(4\beta)/2\\&
-(C_\ell^{EE, \mathrm{N}} - C_\ell^{BB, \mathrm{N}})\sin(4\alpha )/2\\&
+C_\ell^{EB, \mathrm{fg}}+ C_\ell^{EB, \mathrm{N}} \cos(4\alpha ) + C_\ell^{EB,\mathrm{CMB}}\cos(4\beta)\\&
+( C_\ell^{E^\mathrm{fg}B^\mathrm{CMB}} + C_\ell^{E^\mathrm{CMB}B^\mathrm{fg}} ) \cos( 2\beta)
+ (C_\ell^{E^\mathrm{fg}E^\mathrm{CMB}} - C_\ell^{B^\mathrm{fg}B^\mathrm{CMB}})\sin(2\beta)\\&
+ (C_\ell^{E^\mathrm{fg}B^\mathrm{N}} + C_\ell^{E^\mathrm{N} B^\mathrm{fg} } )  \cos(2\alpha )
-(C_\ell^{E^\mathrm{fg}E^\mathrm{N} } - C_\ell^{ B^\mathrm{fg}B^\mathrm{N} } )\sin(2\alpha)\\&
+( C_\ell^{E^\mathrm{CMB}B^\mathrm{N} } + C_\ell^{ E^\mathrm{N} B^\mathrm{CMB} } )\cos(2\alpha -2\beta)\\&
-( C_\ell^{E^\mathrm{CMB}E^\mathrm{N} } - C_\ell^{ B^\mathrm{CMB}B^\mathrm{N} } )\sin(2\alpha -2\beta) \,.
\end{split}
\end{equation}
If we divide throughout by $\cos(4\alpha)$, then among these terms the following do not vanish upon ensemble average: 
\begin{equation}\label{eq:GeneralRotationFitting}
\begin{split}
\langle C_\ell^{EB,\mathrm{o}}\rangle =&
\frac{\tan(4\alpha ) }{2}
\left(\langle C_\ell^{EE,\mathrm{o}}\rangle-\langle C_\ell^{BB,\mathrm{o}}\rangle\right)
+
\frac{\sin(4\beta)}{2\cos(4\alpha)}\left(
\langle C_\ell^{EE,\mathrm{CMB}}\rangle - \langle C_\ell^{BB,\mathrm{CMB}}\rangle
\right)\\
&+\frac1{\cos(4\alpha)}\langle C_\ell^{EB,\mathrm{fg}}\rangle
+\frac{\cos(4\beta)}{\cos(4\alpha)}\langle C_\ell^{EB,\mathrm{CMB}}\rangle\,.
\end{split}
\end{equation} 
The last term, $\langle C_\ell^{EB,\mathrm{CMB}}\rangle$, is the intrinsic $EB$ correlation at the last scattering surface.
This term could arise from,
e.g., chiral gravitational waves~\cite{Lue:1998mq,Saito:2007kt,Sorbo:2011rz}, anisotropic inflation~\cite{Watanabe:2010bu}, etc.
We can measure this signal if its multipole dependence is sufficiently different from that of $\langle C_\ell^{EE,\mathrm{CMB}}\rangle - \langle C_\ell^{BB,\mathrm{CMB}}\rangle$~\cite{Thorne:2017jft}.
Therefore, we shall focus on the cosmic birefringence term and ignore $\langle C_\ell^{EB,\mathrm{CMB}}\rangle$ throughout this paper without loss of generality. 

For the moment, we shall also assume that the ensemble average of the intrinsic $EB$ correlation of the foreground emission vanishes over the full sky, i.e., $\langle C_\ell^{EB,\mathrm{fg}}\rangle=0$.
Of course, a non-zero $EB$ arises from a statistical fluctuation in one realisation of our sky.
The question is whether a $EB$ correlation is statistically significant compared to the cosmic-variance uncertainty.
The current data show no evidence for non-zero $EB$ correlation from the foreground emission~\cite{Adam:2014bub,planckdust:2018};
however, it is still possible that more sensitive future experiments may find a statistically-significant $EB$ correlation in the foreground. We
shall show how to deal with this term in Sect.~\ref{sec:conclusions}.

\subsection{Likelihood analysis}
We determine $\alpha$ and $\beta$ by fitting $C_\ell^{EB, \mathrm{o}}$ with $C_\ell^{EE,\mathrm{o}}- C_\ell^{BB, \mathrm{o}}$ and $C_\ell^{EE,\mathrm{CMB}}- C_\ell^{BB, \mathrm{CMB}}$ using Eq.~(\ref{eq:GeneralRotationFitting}).
To this end, we use a likelihood analysis. The log-likelihood function is given by
\begin{equation}\label{eq:LikelihoodGeneral}
-2\ln \mathcal{L}
= \sum_{\ell=2}^{\ell_\mathrm{max}}
\frac{
	\left[
		C_\ell^{EB,\mathrm{o} } 
	 	- \frac{\tan(4\alpha) }{ 2 }  \left( C_\ell^{EE,\mathrm{o} } - C_\ell^{ BB,\mathrm{o} } \right) 	
		-  \frac{ \sin(4\beta) }{ 2\cos( 4\alpha ) }
	\left(
		C_{\ell}^{EE,\mathrm{CMB} } - C_{\ell}^{BB,\mathrm{CMB} } 
	\right)	
	\right]^2
}{
	\Var \left( 
		C_\ell^{ EB, \mathrm{o} } - \frac{\tan(4\alpha) }{ 2 }  \left( C_\ell^{EE,\mathrm{o} } - C_\ell^{ BB,\mathrm{o} } \right) 
	\right)
}.
\end{equation}
The expression of variance in the likelihood is given in Appendix~\ref{sec:variance}.
As we do not know the $C_{\ell}^{EE,\mathrm{CMB} }$ and $C_{\ell}^{BB,\mathrm{CMB} }$ realised in our sky, we replace them by the best-fitting $\Lambda$CDM theoretical power spectra multiplied by the beam transfer functions, i.e., $C_{\ell}^{EE,\mathrm{CMB,th} }b_\ell^2$ and $C_{\ell}^{BB,\mathrm{CMB,th} }b_\ell^2$, respectively. 
In principle, we could marginalise the likelihood over the difference between $C_{\ell}^{XX,\mathrm{CMB} }$ and $C_{\ell}^{XX,\mathrm{CMB,th}}$ by adding $\left[(C_{\ell}^{XX,\mathrm{CMB}} - C_{\ell}^{XX,\mathrm{CMB,th}}b_\ell^2)^2(2\ell+1)/(2C_{\ell}^{XX,\mathrm{CMB,th}}b_\ell^2)^2\right]$ with $X=E$ or $B$.
In practice, we find that marginalisation has little effect on the uncertainty of $\alpha$ and $\beta$.

We minimise Eq.~(\ref{eq:LikelihoodGeneral})
with respect to $\alpha$ and $\beta$,
given $C_\ell^{ EB, \mathrm{o}}$, $\left(C^{EE,\mathrm{o}}-C^{BB,\mathrm{o}}\right)$, $C_\ell^{EE,\mathrm{CMB,th}}$, and $C_\ell^{BB,\mathrm{CMB,th}}$. 
Assuming flat priors on $\alpha$ and $\beta$, the likelihood gives the posterior distribution of $\alpha$ and $\beta$.

\section{Results}\label{sec:results}
\subsection{Sky simulations}
To validate our methodology, we use the ``PySM" package~\cite{Thorne:2016ifb} to produce realistic simulations of the microwave sky,
with an experimental specification similar to the future satellite mission LiteBIRD~\cite{Hazumi2019}.
Specifically, we include the polarised Galactic foreground emission (``s1'' synchrotron model and ``d1'' dust emission, as described in ref.~\cite{Thorne:2016ifb}),
a CMB map generated from the power spectra based on CAMB~\cite{Lewis:2000},
and white noise with standard deviation given by 
$\sigma_\mathrm{N} = (\pi/10800)(w_{\rm p}^{-1/2}/\mu{\rm K~arcmin})~\mu \mathrm{K~str^{-1/2}} $~\cite{Katayama:2011eh}
with $w_{\rm p}^{-1/2}$ given in the ``Polarisation Sensitivity'' column of Table~\ref{tab:LBspec}.

\begin{table}
	\centering
	\caption{Polarisation sensitivity and beam size of the LiteBIRD telescopes~\cite{Hazumi2019}}\label{tab:LBspec}
	\begin{tabular}{c c c }
		\toprule
		Frequency (GHz) & Polarisation Sensitivity ($\mathrm{\mu K^{'}}$) & Beam Size in FWHM (arcmin) \\ 
		\midrule
		40 & 37.5 & 69 \\
		50 & 24.0 & 56 \\
		60 & 19.9 & 48 	\\
		68 & 16.2 & 43 \\
		78 & 13.5 & 39 \\
		89 & 11.7 & 35 \\
		100 & 9.2 & 29 \\
		119 & 7.6 & 25 \\
		140 & 5.9 & 23 \\
		166 & 6.5 & 21 \\
		195 & 5.8 & 20 \\
		235 & 7.7 & 19 \\
		280 & 13.2 & 24 \\
		337 & 19.5 & 20\\
		402 & 37.5 & 17 \\
		\bottomrule
	\end{tabular}
\end{table}

We use the HEALPix package~\cite{Gorski:2004by} to generate maps with the resolution parameter $N_{\rm side}=512$.
To incorporate beam smearing,
the spherical harmonics coefficients of the CMB and foreground maps at each frequency are multiplied by a beam transfer function, $b_\ell$,
for which we assume a Gaussian beam with  full-width-at-half-maximum (FWHM) given in the third column of Table~\ref{tab:LBspec}.
We calculate the power spectra up to $\ell_\mathrm{max}=2N_\mathrm{side}=1024$.

All of the results reported below will be derived from one realisation of the CMB,
which is common to all frequencies,
and one realisation of noise generated at each frequency.
We do not generate many different realisations,
since the foreground emission is always common to those realisations and thus the uncertainty derived from the ensemble will miss the cosmic-variance contribution of the foreground emission.
In our likelihood (Eq.~\ref{eq:LikelihoodGeneral}),
this foreground cosmic-variance contribution is included in the variance term in the denominator,
which is derived in Appendix~\ref{sec:variance}.
We have checked that the foreground $EB$ correlation in the PySM simulations is consistent with zero within the cosmic variance.

\subsection{Uncertainties on $\alpha$ and $\beta$}
\begin{figure}
	\centering
	\includegraphics[width=1\linewidth]{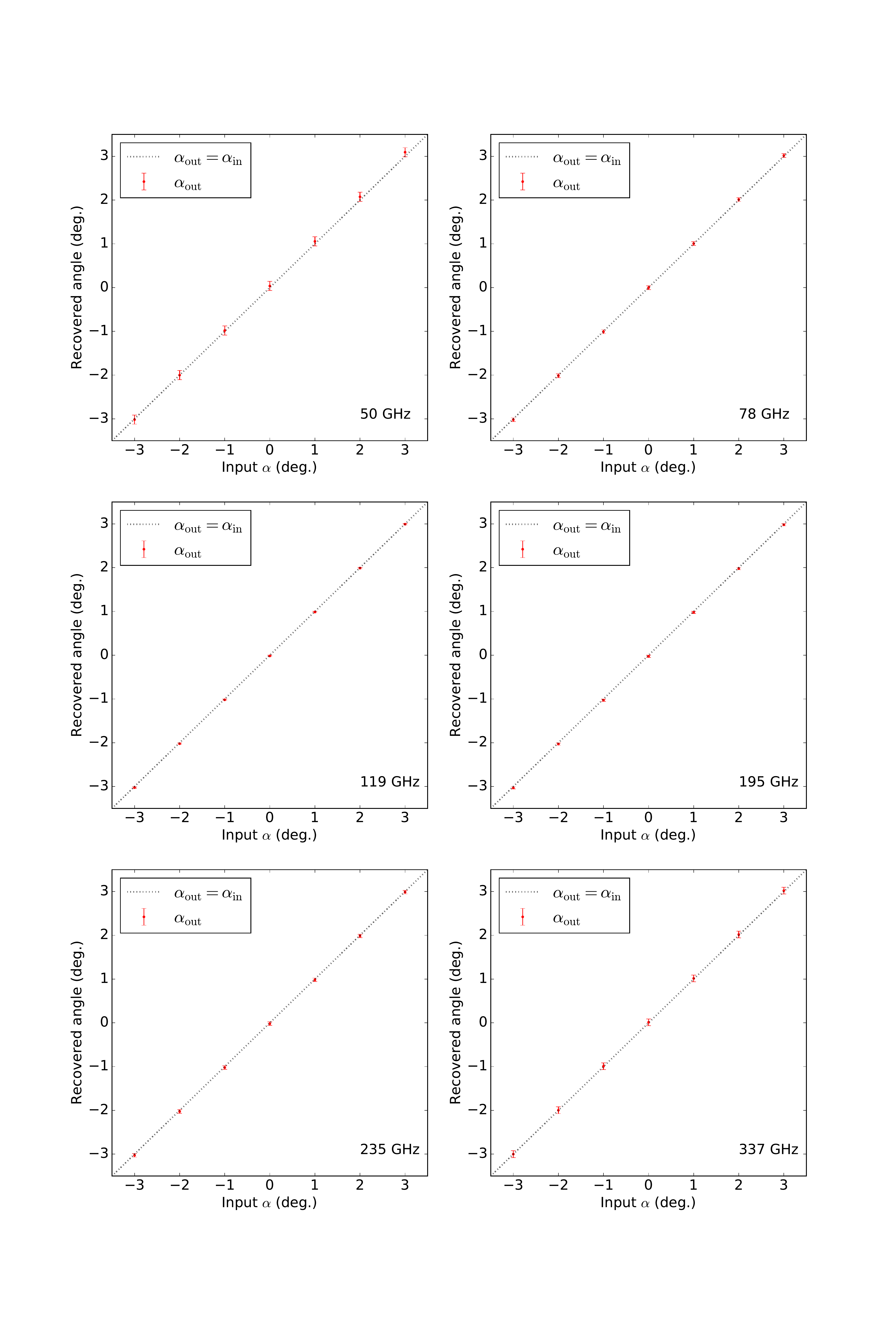}
    \vspace{-25mm}
	\caption{
	Recovery of the miscalibration angle $\alpha$ in the absence of the cosmic birefringence (i.e., with $\beta=0$).
	Each panel shows the recovered values of $\alpha_\mathrm{out}$ (red dots with error bars, in units of degrees) against the input values  $\alpha_\mathrm{in}$ for 6 frequency bands out of 15 specified in Table~\ref{tab:LBspec}.
	}
	\label{fig:NormalRecoveryOfVariedAlpha}
\end{figure}
\begin{figure}
	\centering
	\includegraphics[width=1\linewidth]{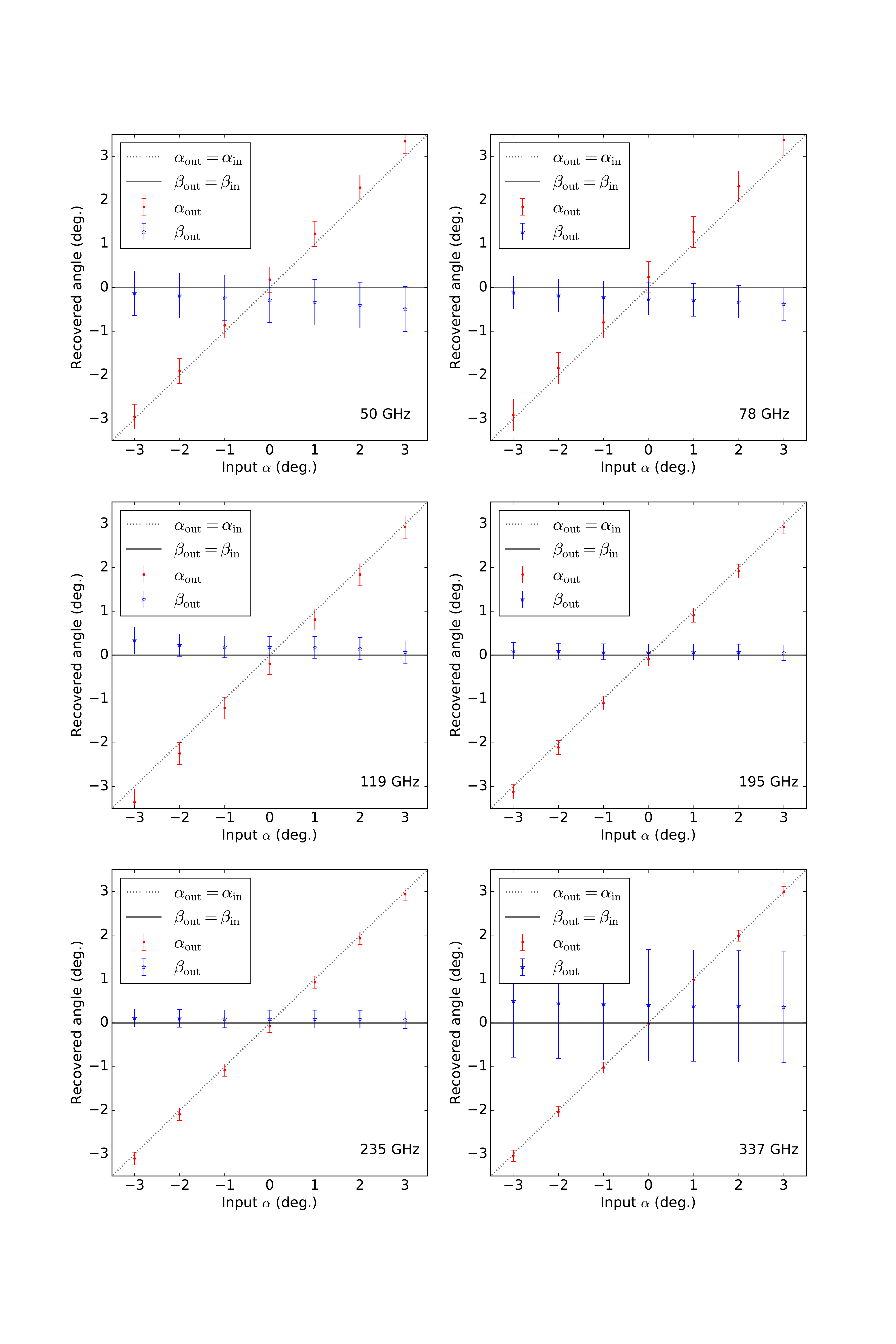}
\vspace{-25mm}
	\caption{
	Simultaneous determination of $\alpha$ and $\beta$ with the input values $\alpha_\mathrm{in}$ varied from $-3^\circ$ to $3^\circ$ and $\beta_\mathrm{in}=0$.
	Each panel shows the recovered values of $\alpha_\mathrm{out}$ (red dots with error bars, in units of degrees) against $\alpha_\mathrm{in}$.
	The blue stars with error bars show the recovered values of $\beta_\mathrm{out}$ at each $\alpha_\mathrm{in}$.
	}
	\label{fig:RecoveryOfVariedAlpha}
\end{figure}
\begin{figure}
	\centering
	\includegraphics[width=1\linewidth]{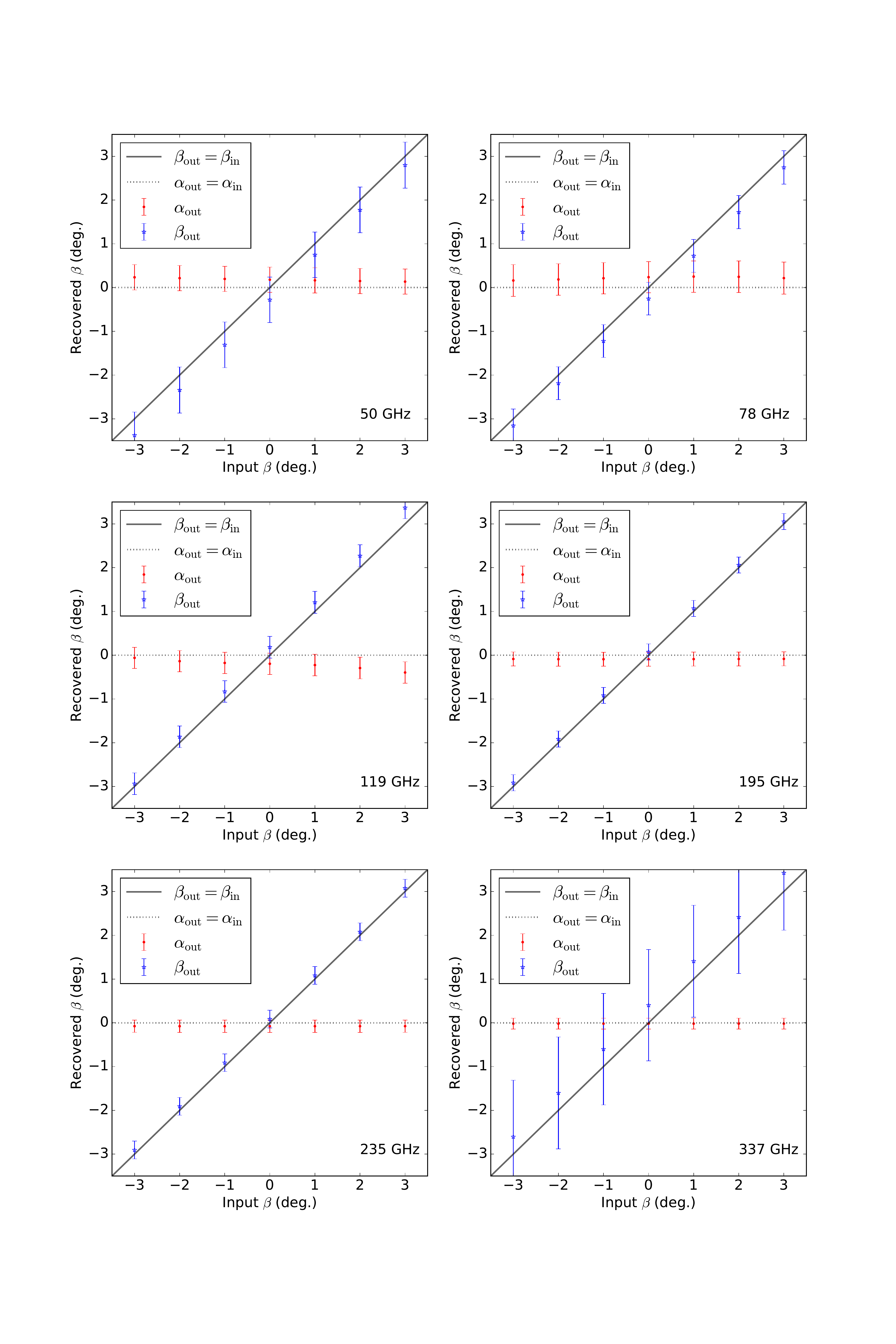}	
	\vspace{-25mm}
	\caption{
	Simultaneous determination of $\alpha$ and $\beta$ with the input values $\alpha_\mathrm{in}=0$ and $\beta_\mathrm{in}$ varied from $-3^\circ$ to $3^\circ$.
	Each panel shows the recovered values of $\beta_\mathrm{out}$ (blue stars with error bars, in units of degrees) against $\beta_\mathrm{in}$.
	The red dots with error bars show the recovered values of $\alpha_\mathrm{out}$ at each $\beta_\mathrm{in}$.
	}
	\label{fig:RecoveryOfVariedBeta}
\end{figure}

First, we report the results with no cosmic birefringence by setting $\beta=0$ in the simulation and fitting only $\alpha$ in the likelihood\footnote{
When the foreground and the angle miscalibration are ignored,
we can interpret $\alpha$ as the birefringence angle. See, e.g., refs.~\cite{Aghanim:2016fhp, MOLINARI201665,Gruppuso_2016,PhysRevD.80.043522,Gruppuso:2016nhj,Pogosian:2019jbt} and references therein for forecasts of the capability of future experiments to constrain the birefringence angle in this simplest case.
}
.

In Figure~\ref{fig:NormalRecoveryOfVariedAlpha}, we show that our method recovers correctly the input values of $\alpha_\mathrm{in}$ from $-3^\circ$ to $3^\circ$ in all frequency bands to within the uncertainties.
We show the 1-$\sigma$ uncertainties, $\sigma(\alpha)$, in the second column of Table~\ref{tab:alphabeta} in units of arcminutes.
The smallest uncertainty is $\sigma(\alpha)=1$\,arcmin, which is achieved at 119 and 140~GHz.

\begin{table}
	\centering
	\caption{
	Marginalised 1-$\sigma$ uncertainties on $\alpha$ and $\beta$ from the experimental specifications given in Table~\ref{tab:LBspec}. The input values are $\alpha_\mathrm{in}=0$ and $\beta_\mathrm{in}=0$
	}\label{tab:alphabeta}
	\begin{tabular}{c|c|cc}
		\toprule
		\multirow{2}{*}{Frequency (GHz)} & $\alpha$ only case (arcmin) & \multicolumn{2}{c}{$\alpha$ and $\beta$ (arcmin)}  \\
		\cmidrule{2-4}
		&$\sigma(\alpha)$ & $\sigma(\alpha)$& $\sigma(\beta)$\\		
		\midrule
		40  &  8.3 &     16 &    70 \\
		50  &  6.3 &     17 &    31 \\
		60  &  4.9 &     20 &    25 \\
		68  &  3.6 &     22 &    24 \\
		78  &  2.6 &     21 &    22 \\
		89  &  1.9 &     19 &    20 \\
		100 &  1.2 &     17 &    17 \\
		119 &  1.0 &     15 &    15 \\
		140 &  1.0 &     12 &    13 \\
		166 &  1.2 &     11 &    12 \\
		195 &  1.5 &     9.6 &    11 \\
		235 &  2.4 &     8.4 &    12 \\
		280 &  4.6 &     8.3 &    26 \\
		337 &  4.7 &     7.5 &    76 \\
		402 &  4.4 &     7.0 &   $4.1 \times 10^2$ \\
		\bottomrule
	\end{tabular}
\end{table}

Next, we determine $\alpha$ and $\beta$ simultaneously.
We show two representative results:
(1) with the input miscalibration angles, $\alpha_\mathrm{in}$, varied from $-3^\circ$ to $ 3^\circ$ while $\beta_\mathrm{in}=0$ (Figure~\ref{fig:RecoveryOfVariedAlpha});
and (2) with the input birefringence angles, $\beta_\mathrm{in}$, varied from $-3^\circ$ to $3^\circ$ while $\alpha_\mathrm{in}=0$ (Figure~\ref{fig:RecoveryOfVariedBeta}).
We find that our method recovers correctly the input values of $\alpha_\mathrm{in}$ and $\beta_\mathrm{in}$ in all frequency bands to within the uncertainties. 

We give the marginalised 1-$\sigma$ uncertainties, $\sigma(\alpha)$ and $\sigma(\beta)$, 
in the third and fourth columns of Table~\ref{tab:alphabeta} in units of arcminutes.
The smallest $\sigma(\beta)$ is $11$\,arcmin at $195$\,GHz (where $\sigma(\alpha)$ is $9.6$\,arcmin).
Simultaneous determination of $\alpha$ and $\beta$ increases the uncertainties on $\alpha$ significantly at all frequency bands.

\begin{figure}
\centering
\begin{tabular}{cc}
\begin{minipage}[t]{0.5\linewidth}
\centering
\includegraphics[width=\linewidth]{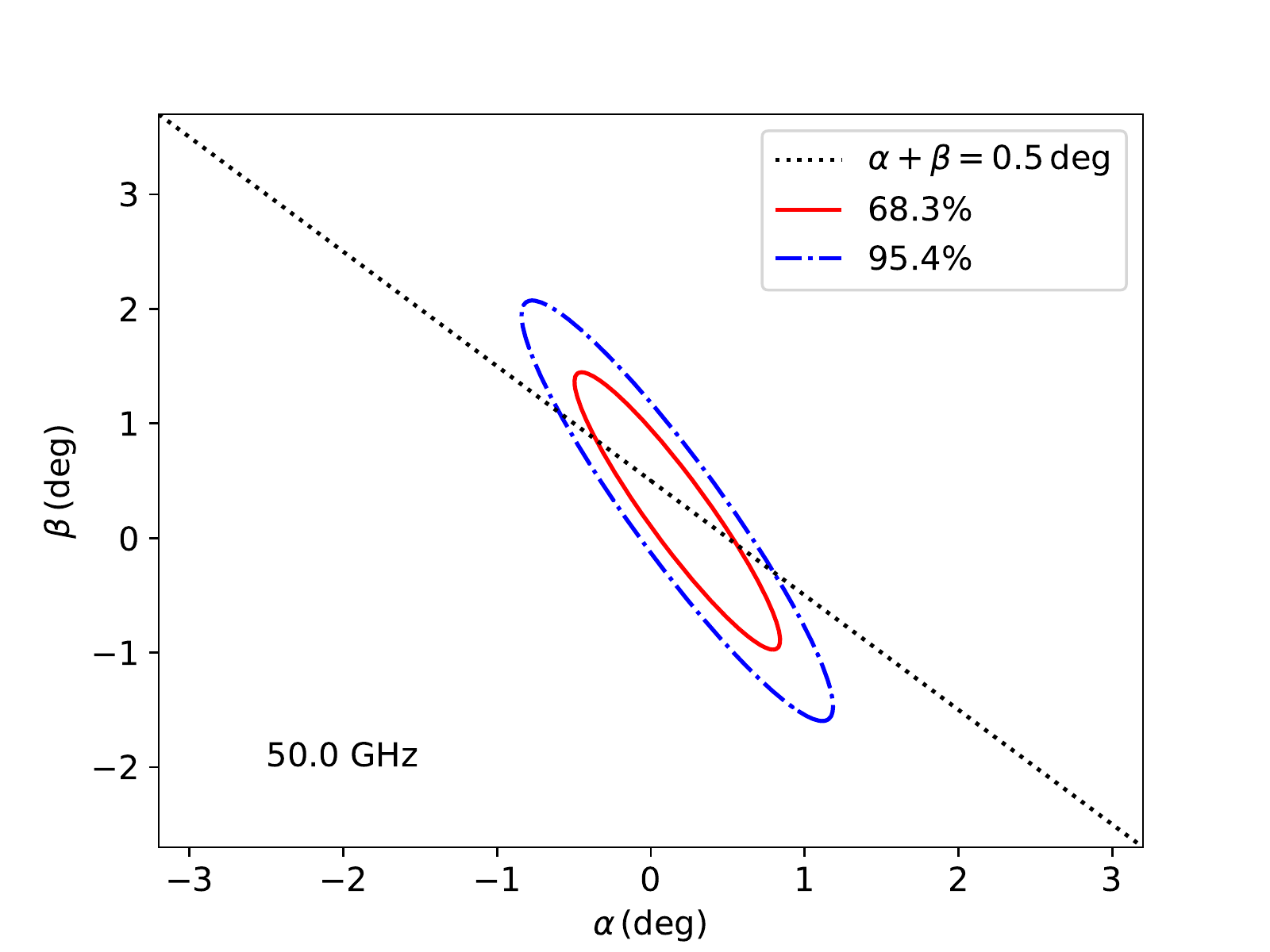}
\end{minipage}
&

\begin{minipage}[t]{0.5\linewidth}
\centering
\includegraphics[width=\linewidth]{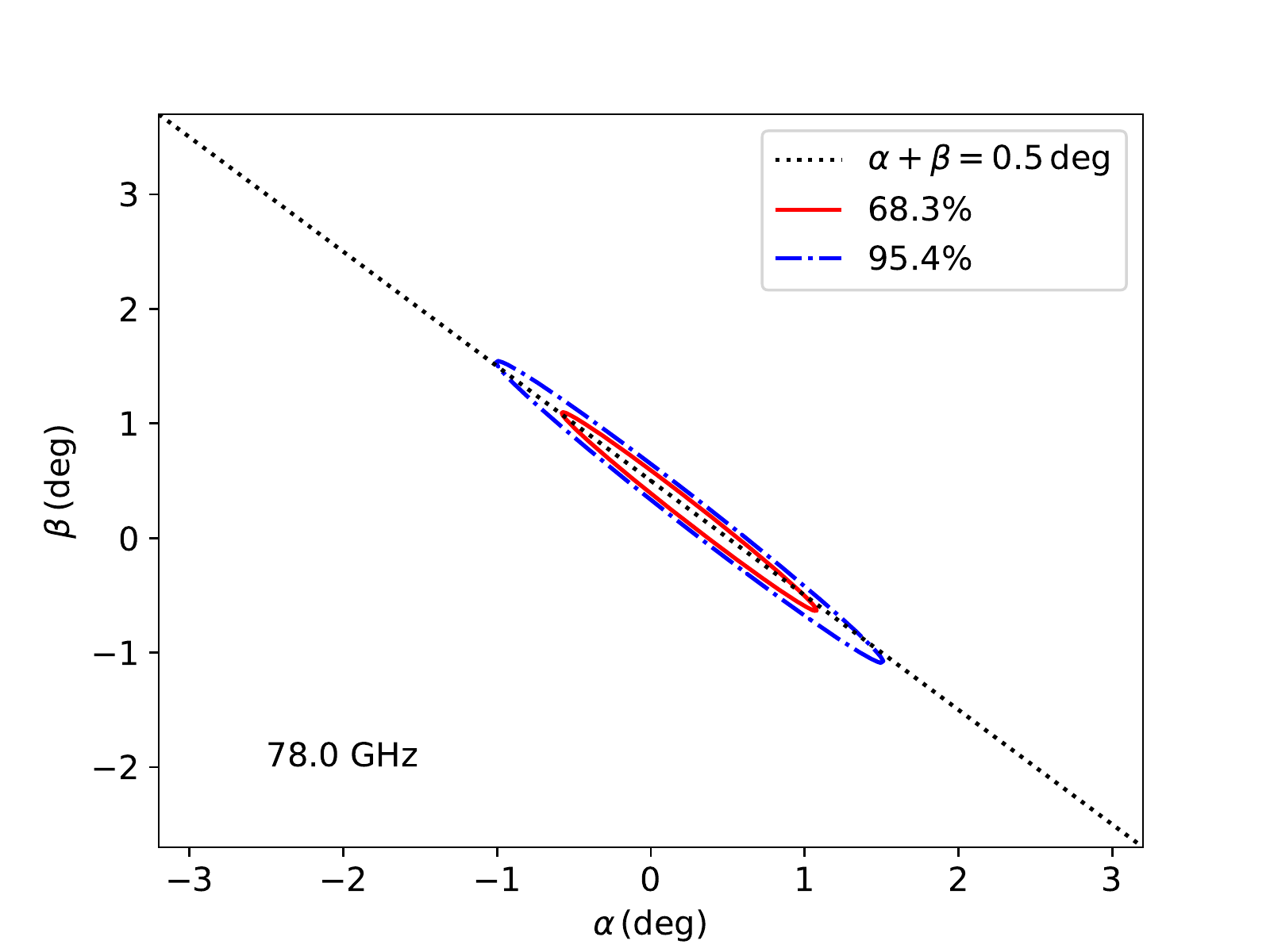}
\end{minipage}
\\

\begin{minipage}[t]{0.5\linewidth}
\centering
\includegraphics[width=\linewidth]{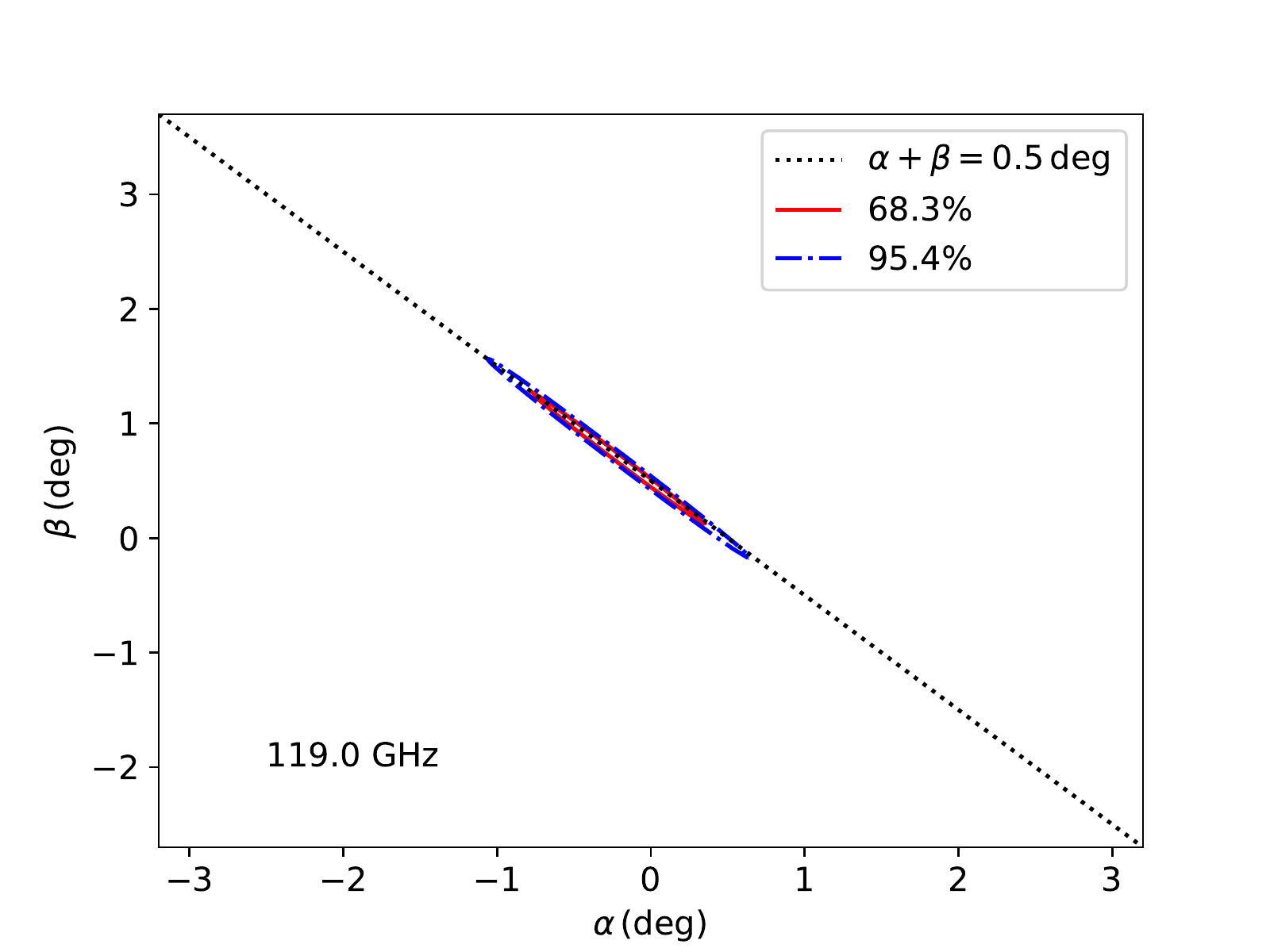}
\end{minipage}
&

\begin{minipage}[t]{0.5\linewidth}
\centering
\includegraphics[width=\linewidth]{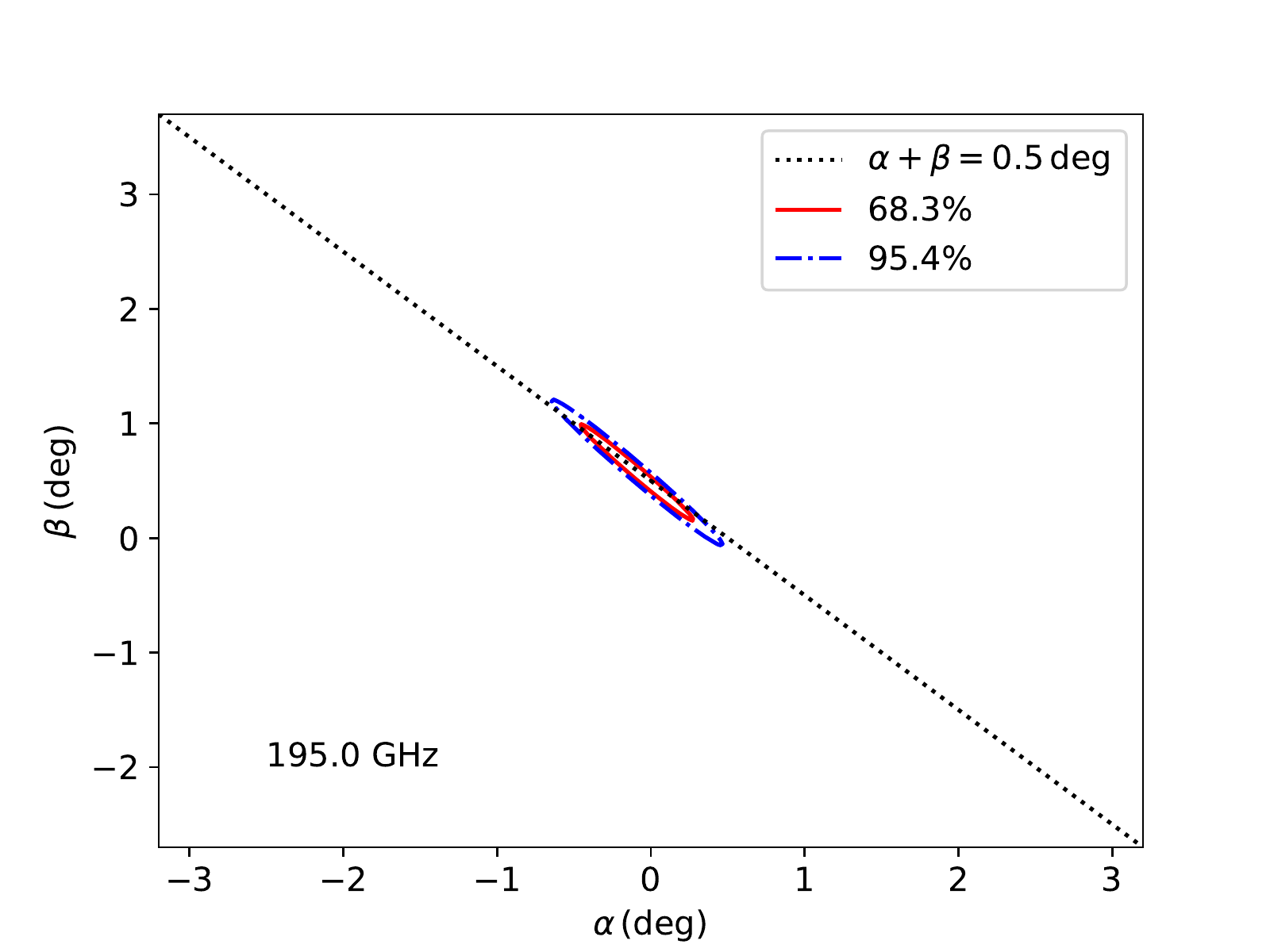}
\end{minipage}

\\

\begin{minipage}[t]{0.5\linewidth}
\centering
\includegraphics[width=\linewidth]{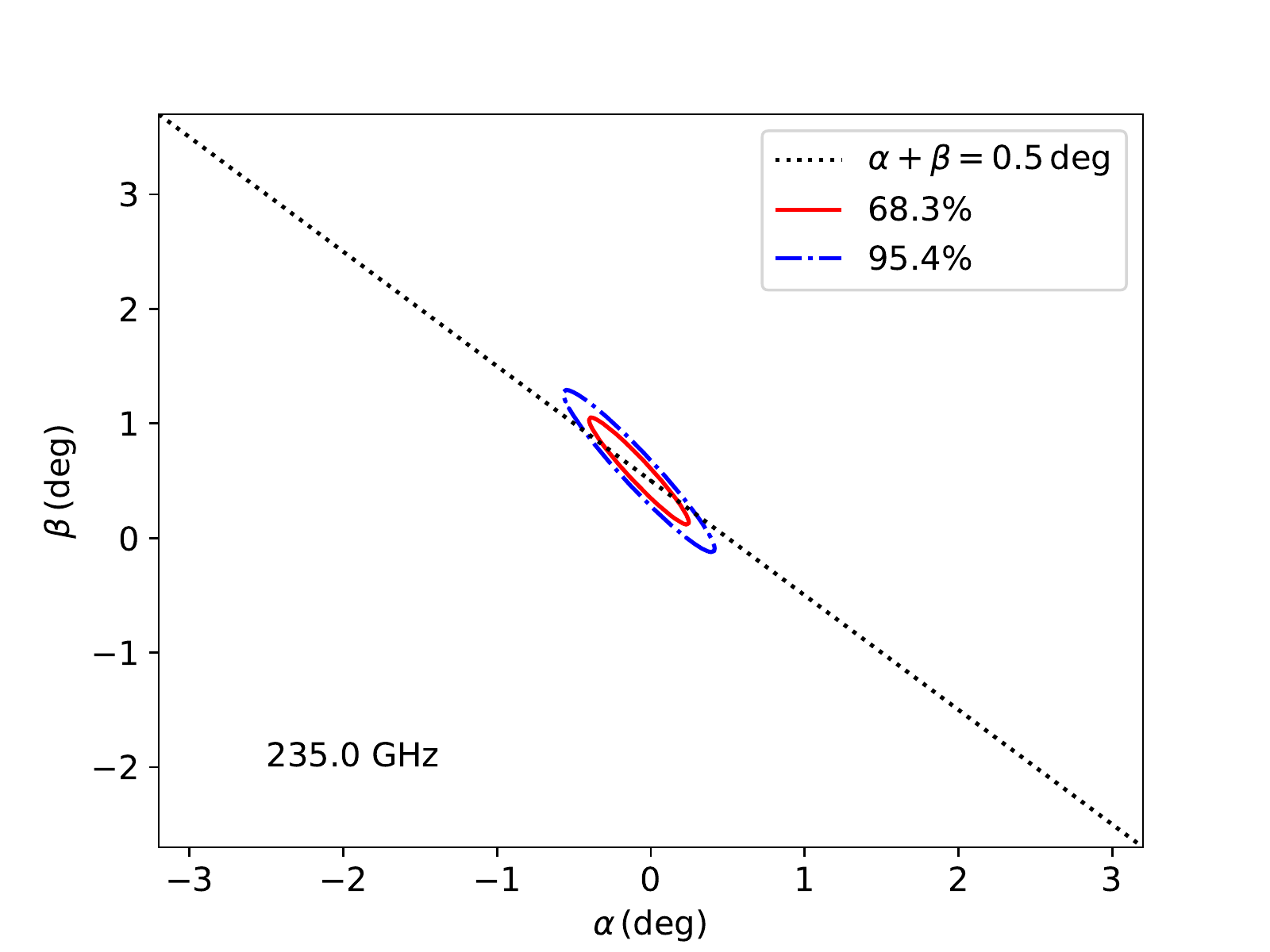}
\end{minipage}

&
\begin{minipage}[t]{0.5\linewidth}
\centering
\includegraphics[width=\linewidth]{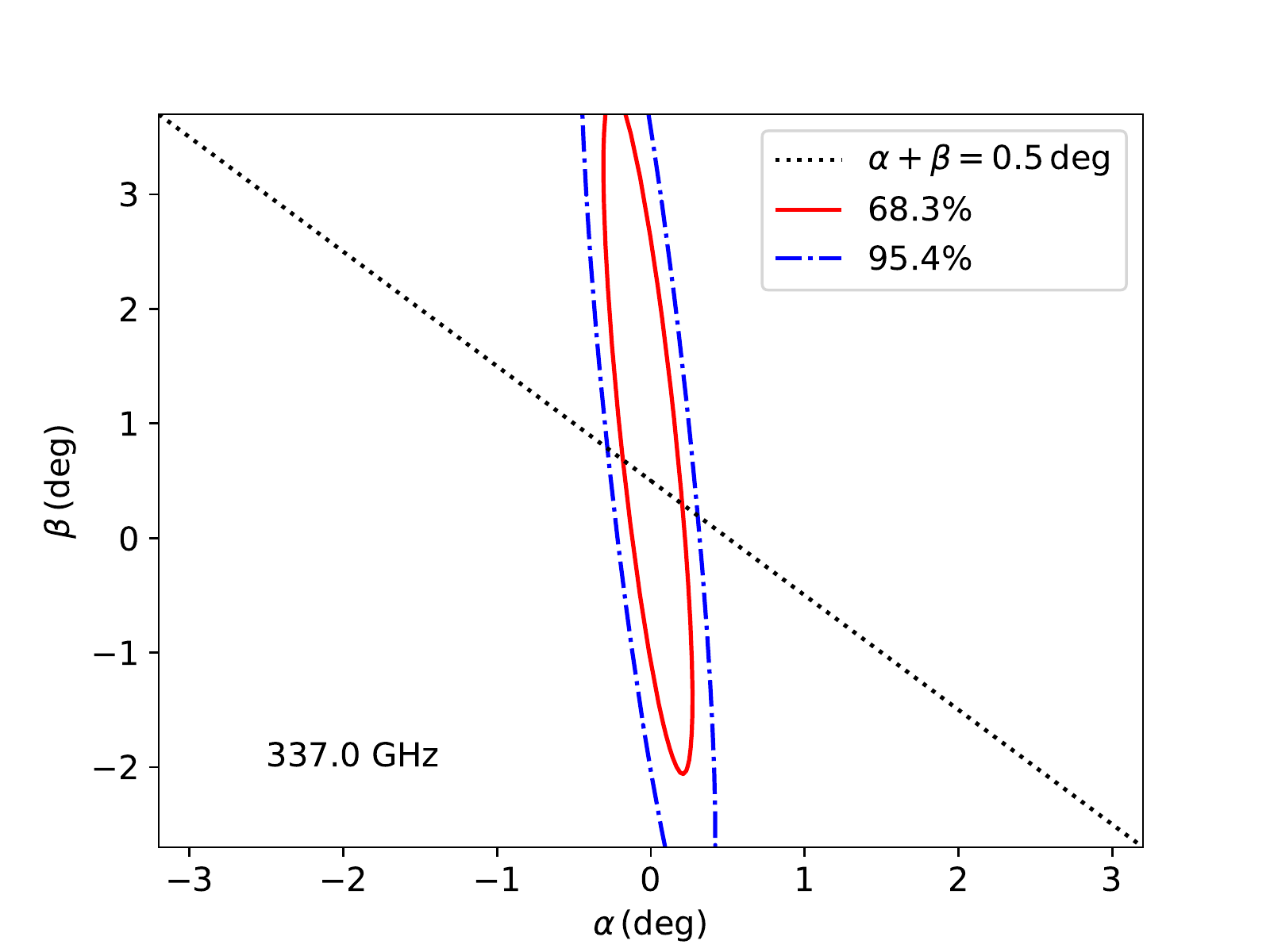}
\end{minipage}
	\end{tabular}
	\caption{
	Joint constraints on $\alpha$ (horizontal axis) and $\beta$ (vertical axis) in units of degrees, from the experimental specifications given in Table~\ref{tab:LBspec}. We show only 6 frequency bands out of 15. 
	The contours show $\Delta(-2\ln\mathcal{L})=2.30$ (68.3\%~CL) and 6.17 (95.4\%~CL). The input values are  $\alpha_\mathrm{in}=0$ and $\beta_\mathrm{in}=0.5^\circ$.
	The black dotted lines show $\alpha+\beta=0.5^\circ$.
	}
	\label{fig:Contour}
\end{figure}

The angles $\alpha$ and $\beta$ are expected to be correlated in any fit. 
When the power spectrum data are dominated by the CMB,
we can only determine a linear combination $\alpha+\beta$.
On the other hand, when the data are dominated by the Galactic foreground emission,
we can only determine $\alpha$.
Therefore, the foreground helps to break degeneracy between $\alpha$ and $\beta$,
allowing us to determine them simultaneously.
We show this in Figure~\ref{fig:Contour}.
The cosmic birefringence angle $\beta$ is poorly constrained in the foreground-dominated frequency bands (lowest and highest frequencies) relative to the miscalibration angle $\alpha$
,
whereas the degeneracy given by $\alpha+\beta$ (dotted lines) is broken in between.
Since $\alpha+\beta$ is constrained tightly by the CMB, adding the foreground yields similar uncertainties on both $\alpha$ and $\beta$.
In other words, the accuracy of $\beta$ is determined by the accuracy of $\alpha$ provided by the foreground emission.

This method can be applied not only to experiments focused on the CMB, but also to calibrate polarisation angles of other experiments focused on the foreground emission,
such as those observing at lower and higher frequencies.
When applied to low frequency data,
it is straightforward to extend our method to incorporate the effect of Faraday rotation on average by rotating the foreground polarisation angle by an additional angle $\gamma\propto \nu^{-2}$, i.e., $\alpha\to\alpha+\gamma$. Accurate subtraction of the Faraday rotation requires estimation of rotation angles per pixel rather than the full-sky average (see, e.g., refs.~\cite{Renzi:2018dbq,Pogosian:2019jbt} and references therein).

Finally, it is possible to combine all frequency bands to obtain the best estimate of $\beta$ with the smallest uncertainty.
To perform such an analysis, however, we must take into account the covariance between different frequency bands.
While the instrumental noise is expected to be uncorrelated to a good approximation, both the CMB and foreground emission are highly correlated across different frequency bands.
We leave the computation of the full likelihood combining all frequency bands to future work.

\section{Discussion and Conclusions}\label{sec:conclusions}
In this paper, we have shown that it is possible to determine the cosmic birefringence and miscalibrated polarisation angles simultaneously from the observed $EB$ cross-correlation power spectrum,
contrary to what has been usually assumed in the literature.
The idea behind our method is simple: the miscalibration angle $\alpha$ affects both CMB and the Galactic foreground emission,
whereas the cosmic birefringence angle $\beta$ affects only CMB. 

The key observation is that the $EB$ correlation induced by the angle miscalibration is related to the difference between the \textit{observed} $EE$ and $BB$ power spectra on the sky, as shown by ref.~\cite{Zhao:2015mqa}.
We can then use accurate knowledge of the CMB polarisation power spectra from the best-fitting $\Lambda$CDM model to separately constrain $\alpha$ and $\beta$.
To this end we have derived the likelihood function given in Eq.~(\ref{eq:LikelihoodGeneral}),
which yields the posterior distribution of $\alpha$ and $\beta$ given the observational data and the best-fitting CMB model.

Applying our method to simulated maps including realistic foreground emission~\cite{Thorne:2016ifb} and CMB as well as instrumental noise and beam smearing similar to the future CMB mission LiteBIRD~\cite{Hazumi2019},
we find that the method successfully recovers the input values of $\alpha$ and $\beta$ simultaneously,
with the minimum uncertainty on the cosmic birefringence being $\sigma(\beta)=11$\,arcmin at 195\,GHz.
Therefore, CMB experiments are capable of constraining parity-violating physics even when we use the $EB$ correlation to self-calibrate polarisation angles as proposed by ref.~\cite{Keating:2012ge}.

So far we have assumed that the intrinsic $EB$ correlation of the foreground emission vanishes when measured over the full sky, i.e.,
$\langle C_\ell^{EB,\mathrm{fg}}\rangle=0$ in Eq.~(\ref{eq:GeneralRotationFitting}).
Let us now address the impact of a possible $EB$ signal from the foreground.
To model this, one could write $\langle C_\ell^{EB,\mathrm{fg}}\rangle = f_c \sqrt{\langle C_\ell^{EE,\mathrm{fg}}\rangle\langle C_\ell^{BB,\mathrm{fg}}}\rangle$ as in ref.~\cite{Abitbol:2015epq}.
As the foreground $EE$ and $BB$ power spectra are similar~\cite{planckdust:2018},
let us write the $BB$ spectrum as $\langle C_\ell^{BB,\mathrm{fg}}\rangle =\xi\langle C_\ell^{EE,\mathrm{fg}}\rangle$,
where $\xi$ is for example equal to $\langle A_{BB}/A_{EE}\rangle\approx 0.5$ for thermal dust emission~\cite{planckdust:2018}.
Then we have
\begin{equation}
    \langle C_\ell^{EB,\mathrm{fg}}\rangle = \frac{f_c\sqrt{\xi}}{1-\xi}\left(\langle C_\ell^{EE,\mathrm{fg}}\rangle-\langle C_\ell^{BB,\mathrm{fg}}\rangle\right).
\end{equation}
This result can be put into the same form as in the angle miscalibration,
if we introduce a new angle $\gamma$ and write $\sin(4\gamma)/2 = f_c\sqrt{\xi}/(1-\xi)$, with
$0 \leq 2f_c\sqrt{\xi}/(1-\xi) \leq 1$.
Therefore, the foreground emission is now rotated by $\alpha(\nu)+\gamma(\nu)$, whereas the CMB is rotated by $\alpha(\nu)+\beta$.
Here, $\nu$ denotes an observing frequency band.
We then use the difference between the multipole dependence of the foreground and the CMB to determine $\beta-\gamma(\nu)$;
thus, we need to give up measuring $\beta$
if we use only one frequency band. 
Fortunately, as the cosmic birefringence effect is independent of frequency,
we can distinguish between $\beta$ and $\gamma(\nu)$ using multi-frequency data.

While we have calculated the expected uncertainties on $\alpha$ and $\beta$ from the future satellite mission LiteBIRD,
we can use the same formalism to calculate those from ground-based and balloon-borne experiments.
This new framework allows us to enhance our ability to constrain parity-violating physics from the CMB polarisation data.

\section*{Acknowledgment}
We thank Joint Study Group of the LiteBIRD collaboration for useful discussion and feedback on this project, and A. Gruppuso, M. Hazumi, D. Molinari, P. Natoli, and D. Scott for comments on the draft. We acknowledge the use of the MINUIT algorithm via the iminuit Python interface.
This work was supported in part by Japan Society for the Promotion of Science (JSPS) KAKENHI Grant Numbers JP15H05896, JP18K03616, JP16H01543, JP15H05891, and JP15H05890, and JSPS Core-to-Core Program, A.
Advanced Research Networks.
\appendix
\section{Variance in the likelihood}\label{sec:variance}
In this section, we derive the variance of $C_\ell^{EB, \mathrm{o}} - ( C_\ell^{EE, \mathrm{o}} - C_\ell^{BB, \mathrm{o}} ) \tan(4\alpha)/2 $ needed in the denominator of Eq.~(\ref{eq:LikelihoodGeneral}).

Using the definition of the power spectrum $C_\ell^{XY}=(2\ell+1)^{-1}\sum_m X_{\ell,m}Y_{\ell,m}^*$ and assuming Gaussianity,
i.e., $\langle X_{\ell,m}Y_{\ell,m}^*{X'}_{\ell,m'}^*{Y'}_{\ell,m'}\rangle=\langle C_{\ell}^{XY}\rangle\langle C_{\ell}^{X'Y'}\rangle
+\langle C_{\ell}^{XX'}\rangle\delta_{mm'}\langle C_{\ell}^{YY'}\rangle\delta_{mm'}
+\langle C_{\ell}^{XY'}\rangle\delta_{m-m'}\langle C_{\ell}^{X'Y}\rangle\delta_{m-m'}$,
we obtain
\begin{align}
\nonumber
&
\Var \left[C_\ell^{EB, \mathrm{o}}   - ( C_\ell^{EE, \mathrm{o}} - C_\ell^{BB, \mathrm{o}})\tan(4\alpha)/2 \right] \\
\nonumber
&=
\Bigl<
\left[C_\ell^{EB, \mathrm{o}}   - ( C_\ell^{EE, \mathrm{o}} - C_\ell^{BB, \mathrm{o}})\tan(4\alpha)/2 \right]^2
\Bigr> 
- \langle C_\ell^{EB, \mathrm{o}}   - ( C_\ell^{EE, \mathrm{o}} - C_\ell^{BB, \mathrm{o}})\tan(4\alpha)/2\rangle^2\\
\nonumber
&=
\frac{1}{2\ell+1}\langle  C_\ell^{EE} \rangle \langle  C_\ell^{BB} \rangle
+\frac{\tan^2(4\alpha)}{4}\frac{2}{2\ell+1}\left(
\langle C_\ell^{EE} \rangle^2
+
\langle C_\ell^{BB} \rangle^2
\right) 
\\
&\quad
-\tan(4\alpha)\frac{2}{2\ell+1} \langle  C_\ell^{EB} \rangle
\left(  \langle  C_\ell^{EE} \rangle -  \langle  C_\ell^{BB} \rangle
\right)
+\frac{1}{2\ell+1}
\left(
1 - \tan^2(4\alpha) 
\right)
\langle  C_\ell^{EB} \rangle^2.
\label{eq:varianceformula}
\end{align}
Here, $\langle C_\ell^{XY}\rangle$ is the sum of all terms including the CMB, foregrounds, and instrumental noise. 

Since we make no assumption about the foreground power spectra,
we shall approximate $\langle C_\ell^{XY}\rangle$ with the observed spectra, $C_\ell^{XY,\mathrm{o}}$.
This is a reasonable approximation at high multipoles, where we have enough statistics.
The fitting results are also dominated by the information at high multipoles.
However, we find that the last term in Eq.~(\ref{eq:varianceformula}) causes a problem: as the observed $EB$ spectrum oscillates around zero due to  the smallness of the signal (if any) and the large scatter from cosmic variance, squaring it yields a biased estimate of the variance. 
Fortunately this term makes only a sub-dominant contribution to the total variance;
thus, we shall ignore it from now on. 
The final formula is then
\begin{equation}
\begin{split}
& 
\Var \left[C_\ell^{EB, \mathrm{o}}   - ( C_\ell^{EE, \mathrm{o}} - C_\ell^{BB, \mathrm{o}})\tan(4\alpha)/2 \right] \\
&\approx
\frac{1}{2\ell+1}C_\ell^{EE,\mathrm{o}}C_\ell^{BB,\mathrm{o}}
+\frac{\tan^2(4\alpha)}{4}\frac{2}{2\ell+1}\left[
(C_\ell^{EE,\mathrm{o}})^2
+
(C_\ell^{BB,\mathrm{o}})^2
\right]
\\
&\quad
-\tan(4\alpha)\frac{2}{2\ell+1}C_\ell^{EB,\mathrm{o}}
\left( C_\ell^{EE,\mathrm{o}} - C_\ell^{BB,\mathrm{o}} 
\right).
\label{eq:varianceformulafinal}
\end{split}
\end{equation}

\bibliographystyle{ptephy}
\bibliography{references}
\end{document}